\documentstyle[12pt,aaspp4]{article}

\lefthead{Gonzalez and Laws}
\righthead{HD\,75289}
\slugcomment{Accepted by the Astronomical Journal}
\received{}
\accepted{}
\begin{document}

\title{Parent Stars of Extrasolar Planets V: HD\,75289}

\author{Guillermo Gonzalez and Chris Laws}

\affil{University of Washington, Astronomy Department, Box 351580, Seattle, 
WA 98195}
\authoremail{gonzalez@astro.washington.edu, laws@astro.washington.edu}

\begin{abstract}

The results of a new spectroscopic analysis of HD\,75289, recently reported to 
harbor a Jovian-mass planet, are presented.  From high-resolution, high-S/N 
ratio spectra, we derive [Fe/H] $=+0.28\pm0.05$ for this star, in agreement 
with the spectroscopic study of Gratton et al., published 10 years ago.  In 
addition, we present a re-analysis of the spectra of $\upsilon$\,And and 
$\tau$\,Boo; our new parameters for these two stars are now in better 
agreement with photometrically-derived values and with the recent 
spectroscopic analyses of Fuhrmann et al.  We also report on extended 
abundance analyses of 14\,Her, HD\,187123, HD\,210277, and $\rho^{1}$\,Cnc.

If we include the recent spectroscopic analyses of HD\,217107 by Randich et 
al. and Sadakane et al., who both reported [Fe/H] $\sim +0.30$ for this star, 
we can state that all the "hot-Jupiter" systems studied to date have 
metal-rich parent stars.  We find that the mean [C/Fe] and [Na/Fe] values 
among the stars-with-planets sample are smaller than the corresponding 
quantities among field stars of the same [Fe/H].

\end{abstract}


\section{Introduction}
\label{intro}

In our continuing series on the parent stars of extrasolar planets (Gonzalez 
1997, Paper I; Gonzalez 1998, Paper II; Gonzalez \& Vanture 1998, Paper III; 
and Gonzalez et al. 1999, Paper IV), we have reported on the results of 
our spectroscopic analyses of these stars.  Other similar studies include 
Fuhrmann et al. (1997, 1998) and Sadakane et al. (1999).  The most significant 
finding so far has been the high mean metallicity of these stars, as a group, 
compared to the metallicity distribution of nearby solar-type stars.  
Additional extrasolar planet candidates continue to be announced by planet 
hunting groups using the Doppler method.  Herein, we report on a Local 
Thermodynamic Equilibrium (LTE) abundance analysis of HD\,75289, which was 
announced on 1 February 1999 (Udry et al. 1999) to harbor a low-mass object 
with a 3.5 day nearly-circular orbit.

In addition to the new candidate listed above, we also present new analyses 
of the spectra of $\upsilon$\,And and $\tau$\,Boo, which had been the subject 
of Paper I.  The basic stellar parameters and abundances of these two stars 
were not well-detemined in that study, due to their relatively broad lines, 
which resulted in a short linelist.  We will improve on that study by adding 
several Fe I,II lines carefully chosen to better constrain the solutions.  
Also, we report on extended spectroscopic analyses of the following parent 
stars: HD\,187123, HD\,210277, 14\,Her, and $\rho^{1}$\,Cnc.  These stars 
were discussed in Papers III and IV, but we had not performed a general 
abundance analysis from their spectra (note, we had included $\rho^{1}$\,Cnc 
in Paper II, but that study was superceded by Paper III).  We close with a 
summary of the abundance patterns among stars with planets and compare them 
with those of nearby F and G dwarfs without (known) planets.

\section{Observations}
\label{obs}

High-resolution, high S/N ratio spectra were obtained with the 2dcoude echelle 
spectrograph (described in Tull et al. 1995) at the McDonald observatory 
2.7 m telescope.  This is the same instrument employed in Papers I to III 
(in Paper IV, we analyzed spectra obtained by G. Marcy with HIRES on the 
Keck I).  The spectral resolving power (determined from the FWHM of the Th-Ar 
lines in the comparison lamp spectrum) is about 65,000, and the S/N ratio is 
about 450-500.  The spectral coverage ranges from 3700 to 10,000 \AA, with 
gaps between orders beyond about 5500 \AA.  The data reduction methods are 
the same as those employed in Papers I to III.  Two spectra of HD\,75289 
were obtained, for a total exposure time of 20 minutes.  Spectra of a hot 
star with a high $v \sin i$ value were also obtained in order to divide out 
telluric lines (with the IRAF program, {\it telluric}).

We derived a heliocentric radial velocity of $+9.9 \pm 0.5$ km~s$^{\rm -1}$ 
(formal error) from our spectra of HD\,75289 obtained on HJD $= 2451215.784$; 
this estimate is based on four clean Fe I lines with laboratory wavelengths 
adopted from Gratton et al.'s (1989) study.  Note that while we did not 
observe a radial velocity standard, the stability of the 2dcoude 
spectrograph should result in systematic velocity errors of no more than 
about 0.5 km~s$^{\rm -1}$.  Our velocity estimate differs significantly from 
Gratton et al.'s radial velocity of $+1$ km~s$^{\rm -1}$, which is the mean 
of six observations they made over a two week period.  Based on a comparison 
with other published velocity estimates for this star, they concluded that 
HD\,75289 is a radial velocity variable (with an amplitude of a few 
km~s$^{\rm -1}$).  However, Udry et al. report a systemic velocity 
of 9.26 km~s$^{\rm -1}$, which is consistent with our single velocity 
estimate.  In addition, they obtain a very good fit with a simple 
Keplerian model, implying that the velocity is not variable at the few 
km~s$^{\rm -1}$ level.

\section{Analysis}
\subsection{Spectroscopic analysis}

The present method of analysis is the same as that employed in Paper III 
for $\rho^{\rm 1}$\,Cnc.  Briefly, it makes use of the line analysis code, 
MOOG (Sneden 1973, updated version), the Kurucz (1993) LTE plane 
parallel model atmospheres, and Fe I, II equivalent width (EW) measurements 
to determine the following atmospheric parameters: T$_{\rm eff}$, $\log g$, 
$\xi_{\rm t}$, and [Fe/H], where the symbols have their usual meanings.  
The Fe linelist was put together from Table 1 of Paper I, Table 2 of Paper II, 
and Table 1 of Paper IV.  Lines of elements other than Fe were selected from 
Table 1 of Paper 1 and Table 2 of Paper II.  We have also added additional 
lines to both lists.  Their $gf$-values were calculated from an inverted 
solar analysis using the Kurucz et al. (1984) Solar Flux Atlas or our 
spectrum of Vesta.  The final linelist, along with the EW values, is listed 
in Tables 1 and 2.  The number of lines employed for each star varies 
somewhat due to the variations in intrinsic line width and temperature from 
one star to another; for example, low excitation lines are weaker in hot 
star spectra, and weak lines are more difficult to measure in spectra with 
relatively high $v \sin i$, i.e., $\upsilon$\,And and $\tau$\,Boo, and in 
cool stars with strong-lined spectra due to crowding, i.e., 14\,Her and 
$\rho^{\rm 1}$\,Cnc.

The abundances of Li and Al were determined via comparison of synthesized 
spectra with the observed spectra.  The method employed to determine the Li 
abundance is described in our previous papers.  Our estimate of the Al 
abundance is based on the Al I pair at 6696 and 6698 \AA; they are 
sufficiently close to the Li I line at 6708 \AA\ that we were able to 
determine the Li and Al abundances with the same synthesized spectral 
region.  This is a change from our previous studies, where we had relied 
primarily on the 7835 and 8772 \AA\ pairs.  Unfortunately, the spectrograph 
setup was such that the 6696 and 6698 \AA\ Al pair fell just outside the 
order containing the Li I line in our spectrum of HD\,75289, so we do not 
quote an Al abundance for this star.

The results of the analyses are presented in Tables 2 to 5.  The calculations 
of the uncertainties and the contribution from systematic errors are 
the same as those discussed in Paper III; systematic errors should be 
negligible in the present study, since these stars are similar to the Sun.  
We argued in Paper III that our LTE analysis of $\rho^{\rm 1}$\,Cnc, which is 
about 500 K cooler than the Sun, probably does not suffer from significant 
systematic errors.  As in Paper III, we refrain from quoting formal 
uncertainties in [Fe/H] less than 0.05 dex.

\subsection{Derived parameters}

We have determined the masses and ages in the same way as in Paper II.  
Using the {\it Hipparcos} parallaxes (ESA 1997) and the stellar 
evolutionary isochrones of Schaller et al. (1992) and Schaerer et al. (1993), 
along with our spectroscopic $T_{\rm eff}$ estimates, we have estimated the 
masses and ages for $\upsilon$\,And, $\tau$\,Boo, and HD\,75289.  Due to 
the large parallaxes and hence small distances, neither the Lutz \& Kelker 
(1973) nor extinction corrections were applied.  We list the results in 
Table 4.  The corresponding theoretical surface gravities are: $\log g = 4.10 
\pm 0.04$, $4.25 \pm 0.03$, and $4.33 \pm 0.02$ for $\upsilon$\,And, 
$\tau$\,Boo, and HD\,75289, respectively.  The close agreement between the 
observed and theoretical surface gravities for all three stars supports the 
assumptions that went into the calculation of the stellar evolutionary 
isochrones and the LTE abundance analyses.

\section{Discussion}

\subsection{$\upsilon$\,And and $\tau$\,Boo}

The present spectroscopic analyses of $\upsilon$\,And and $\tau$\,Boo are a
significant improvement over those reported in Paper I, both as evidenced 
by the reduction in the uncertainties of the derived physical parameters 
and the closer agreement with the spectroscopic analyses of Fuhrmann et al. 
(1998) and photometrically-derived parameters.  While our new $T_{\rm eff}$ 
estimates are significantly smaller than those in Paper I - $\upsilon$\,And 
is less by 110 K and $\tau$\,Boo is less by 180 K - the [Fe/H] values are 
similar; this is due to the fact that we derived a complete new set of 
atmosphere parameters for each star, not just a new $T_{\rm eff}$.  As a 
result, the basic conclusions of Gonzalez (1999a), which is a study of the 
chemo-dynamical properties of stars-with-planets, are not altered.  In 
particular, the conclusion that $\tau$\,Boo possesses an anomalously high 
[Fe/H] value for its age and Galactocentric distance still holds.  Finally, 
we note that the value of $\xi_{\rm t}$ we obtain for $\tau$\,Boo is unusually 
small compared to the other hot stars we analyzed; its $\xi_{\rm t}$ value 
should be larger, given its higher $T_{\rm eff}$ value.

\subsection{HD\,75289}

The {\it Bright Star Catalog} (Hoffleit 1982) designates HD\,75289 as 
G0Ia-0:, which is clearly incorrect.  Gratton et al. included HD\,75289 in 
their spectroscopic abundance study of G and K supergiants 
(their work confirmed that HD\,75289 was in fact a metal-rich dwarf and not 
a supergiant).  Of the EW measurements reported in their paper and ours, 
there are 15 spectral lines in common, with an average difference of 
only $+1.5$ m\AA\ between them. They went on to derive the atmospheric 
parameters based on a total of 35 Fe I and 5 Fe II lines, and obtained the 
following results: $T_{\rm eff} = 6000$ K, $\log g = 3.8$, $\xi_{\rm t} = 
1.3$~km~s$^{\rm -1}$, and [A/H] $= 0.2$.  We note that the present work, which 
uses a similar method of analysis and a larger number of Fe lines to better 
constrain these same stellar parameters, is in close agreement with their 
results.  The only exception to this statement is $\log g$, where our derived 
value of 4.47 differs considerably.  The stellar evolutionary $\log g$ value 
tends to support our estimate.

Gonzalez (1999a) compared the [Fe/H] estimates of the parent stars to the 
mean trends of [Fe/H] with age and mean Galactocentric distance, $R_{\rm m}$, 
among field stars.  Among the young stars (age $\le 2$ Gyr), not only is 
$\tau$\,Boo too metal-rich for its value of $R_{\rm m}$, but so is HD\,75289.  
They are both metal-rich relative to the typical field star of the same 
$R_{\rm m}$ by $+0.26$ dex.

Henry et al. (1996) reported a $\log R^{\rm '}_{\rm HK}$ value of $-5.00$ 
for HD\,75289 from a single measurement; we confirm the low chromospheric 
activity level of this star from examination of the Ca II H and K lines in 
our spectra.  This measure places it among the low-activity stars of the 
roughly 800 stars observed by them.  Employing the activity-age relation of 
Donahue (1993)\footnote{As reported in Henry et al. (1996).} we derive an age 
of 5.6 Gyr, nearly a factor of three greater than the age derived from its 
position on the HR diagram.  None of the other parent stars displays such a 
large activity age relative to the evolutionary age\footnote{70\,Vir has an 
evolutionary age nearly four times its activity age, but this discrepancy is 
likely due to its more evolved state than the other parent stars.}.  Further, 
Udry et al. reported a $v \sin i$ value of 4.37 km~s$^{\rm -1}$ - about half 
the $v \sin i$ value of $\upsilon$\,And, itself 3.5 Gyr old according to its 
location on the HR diagram.

Hence, according to its chromospheric activity level and rotation, HD\,75289 
is older than the Sun, while it is younger according to stellar evolution.  A 
possible way out of this dilemma may be to invoke a phenomenon that spun-down 
HD\,75289 faster than is typical for stars of its spectral type.

\subsection{Abundance Trends}

To search for subtle abundance anomalies among the star-with-planets sample, 
we will compare our results to high-quality abundance analyses of the general 
field population.  The best sources of data on abundances of field stars are 
Edvardsson et al. (1993), Tomkin et al. (1997), Feltzing \& Gustafsson 
(1998), and Gustafsson et al. (1999).  All four studies are based on the 
Uppsala Astronomical Observatory group analysis techniques, and, hence, 
should be consistent with each other.  In addition to these, we will make use 
of several studies of Li abundances among field and open cluster stars.  In 
the following, for most elements, we will compare abundances relative to Fe 
(as [X/Fe]), since such a quantity is less sensitive to systematic differences 
among various studies.

Among the elements measured in the stars-with-planets sample, Li has the 
potential to give us the greatest insight into the process of planet 
formation.  Its abundance in a stellar atmosphere is affected by a number of 
physical processes, some of which are possibly related to the presence of 
planets (see discussion in Paper II).  Among the stellar parameters found to 
correlate with Li abundance are $T_{\rm eff}$, age, and metallicity (Pasquini, 
Liu, \& Pallavicini 1994).  What's more, the Li abundance on the surface of 
an F star might be enhanced as a result of the accretion of rocky material 
(see Alexander 1967 and Paper II).  Following Pasquini et al., we have 
derived an equation relating $\log \epsilon(Li)$ to $T_{\rm eff}$, the 
chromospheric emission measure ($R^{\rm '}_{\rm HK}$), and [Fe/H]:

\begin{equation}
{\log \epsilon(Li) = -80.246 + 0.806 [Fe/H] + 0.431 
\log R^{\rm '}_{\rm HK} + 22.436 \log T_{\rm eff}}
\end{equation}

The stars used to calibrate this equation are from Pasquini et al., Favata 
et al. (1997), and Randich et al. (1999) with the $R^{\rm '}_{\rm HK}$ values 
from Henry et al. (1996).  The range of applicability of the parameters are: 
$-0.61 \le$ [Fe/H] $\le +0.22$, $5458 \le T_{\rm eff} \le 6180$ K, 
$-5.24 \le \log R^{\rm '}_{\rm HK} \le -4.34$, and $+0.83 \le \log 
\epsilon(Li) \le +2.92$.  Note, some of our stars are outside the metallicity 
range of equation 1.  All the stars-with-planets but one have observed Li 
abundances less than the values calculated from equation 1 (Figure 2).  
The largest deviation on this plot is $\tau$\,Boo; there are two points 
to note about it: its $T_{\rm eff}$ value is beyond the range for which 
equation 1 is calibrated, and it is within the so-called "Li dip" seen among 
open cluster stars (Balachandran 1995).  It is also important to note that 
many stars of the same temperature, age, and metallicity range as those 
used to determine equation 1 do not have detectable Li; an example of this 
is the large spread in Li among single stars of the same colors in M67 
(Jones, Fischer, \& Soderblom 1999).  The Li abundance of HD\,75289 is 
not unusual compared to the field star sample, but it might be slightly high 
with respect to its evolutionary age.

The study of Gustafsson et al. is probably the most accurate study of C 
abundances among F and G disk to date.  They employed the [C I] line at 8727 
\AA.  The [C/Fe] values display remarkably small scatter about a mean trend 
with respect to [Fe/H] (see Figure 4 of Gustafsson et al.).  In Figure 1 we 
present the [C/Fe] estimates from Gustafsson et al. and Tomkin et al. (who 
did not employ the [C I] line) for field stars as well as the 
stars-with-planets.  A small trend of [C/Fe] with Galactocentric distance 
has been removed from the individual data points (amounting to $-0.015$ dex 
per kpc).  Some of the Tomkin et al. stars and all but one of the 
stars-with-planets fall below the mean trend line; $\tau$\,Boo displays the 
largest negative deviation.  The [C/Fe] estimate for HD\,217107 is from 
Sadakane et al.'s two measurements: the [C I] and 5380 \AA\ lines; the 
estimate for 51\,Peg is the average from Paper II and Tomkin et al.  It is 
always possible that there is a systematic offset between our [C/Fe] estimates 
and those of Gustafsson et al. due to the different lines used, but it is not 
likely to be significant since both studies are differential relative to the 
Sun.  The deviation of the Sun's [C/Fe] value from the mean trend in Figure 
1 is also notable; while it may not seem like a large difference, the error 
bars on the data point corresponding to the Sun are effectively zero, since 
Gustafsson et al.'s study is differential with respect to the Sun (for 
additional discussion on this point see Gustafsson et al. and Gonzalez 1999b).

Feltzing \& Gustafsson examined abundance trends (as [X/Fe] versus [Fe/H]) 
among metal-rich disk stars.  For most elements, there are no significant 
deviations from the solar ratios, but they did find a significant upturn in 
[Na/Fe] for stars with [Fe/H] $> 0.00$, reaching [Na/Fe] $\sim 0.20$ for the 
most metal-rich stars.  Among the stars-with-planets sample, the mean [Na/Fe] 
value is $-0.02$; $\rho^{\rm 1}$\,Cnc, HD\,75289, and HD\,210277 have the 
smallest [Na/Fe].  They used the same two Na I lines we employed in our 
study.  The mean values of [X/Fe] for the other elements among our sample 
stars do not appear to differ significantly from the trends seen among disk 
stars.

The most obvious abundance trend among the stars-with-planets studied so far 
is their high mean metallicity compared to the general field population.  Our 
estimate for the [Fe/H] value of HD\,75289, $+0.28$, is close to the mean 
of the so-called "hot Jupiter" systems.  Another recently 
announced system, HD\,217107, was studied spectroscopically by Randich et al. 
and Sadakane et al., who obtained [Fe/H] $= +0.30$ and $+0.31$, 
respectively.

\section{Conclusions}

The results of our analysis of high-resolution spectra of HD\,75289 confirm 
that it is a metal-rich star, with [Fe/H] $= +0.28$.  Its evolutionary 
age, 2.1 Gyr, is much less than the age derived from its chromospheric 
emission measure, $R^{\rm '}_{\rm HK}$.

Compared, as a group, to nearby F and G dwarfs, the stars-with-planets sample 
display the following peculiarities:

\begin{itemize}
\item
The latest additions to this group, HD\,75289 and HD\,217107, continue the 
trend, first noted in Paper I, that stars-with-planets are metal-rich relative 
to the nearby field star population.
\item
The stars, $\tau$\,Boo, $\rho^{\rm 1}$\,Cnc, 14\,Her, HD\,75289, and 
HD\,217107, are much more metal-rich than F and G dwarfs of similar ages 
and mean Galactocentric distances.
\item
Compared to field stars with detectable Li, stars-with-planets tend to have 
smaller Li abundances when corrected for differences in $T_{\rm eff}$, [Fe/H], 
and $R^{\rm '}_{\rm HK}$.
\item
The [Na/Fe] and [C/Fe] values of stars-with-planets are, on average, smaller 
than the corresponding quantities among field stars of the same [Fe/H].
\end{itemize}

In summary, while the numbers are still small, the data on stars-with-planets 
are beginning to indicate ways in which they differ from the general field 
star population.  These abundance anomalies might be useful in constraining 
future searches for extrasolar planets, and they will be very helpful in 
theoretical studies of planet formation.

\acknowledgements
  
The authors are grateful to David Lambert for obtaining spectra of HD\,75289 
at our request.  Thanks also go to Robert Kurucz for his model atmospheres and 
Chris Sneden for use of his code, MOOG.  This research has made use of the 
Simbad database, operated at CDS, Strasbourg, France.  The research has been 
supported in part by the Kennilworth Fund of the New York Community Trust.

\clearpage

\begin{deluxetable}{lcccc}
\tablecaption{Atomic Data and Equivalent Widths for HD\,75289}
\tablewidth{0pt}
\tablehead{
\colhead{Species} & \colhead{$\lambda_{\rm o}$} & 
\colhead{$\chi_{\rm l}$} & \colhead{$\log gf$} & 
\colhead{EW}\\
\colhead{} & \colhead{(\AA)} & \colhead{(eV)} & \colhead{} & 
\colhead{(m\AA)}
}
\startdata
C I & 5380.32 & 7.68 & $-1.71$ & 39.2\nl
C I & 6587.62 & 8.53 & $-1.08$ & 28.9\nl
C I & 7483.42 & 8.77 & $-1.46$ & 14.5\nl
N I & 7468.27 & 10.33 & $-0.02$ & 10.7\nl
Na I & 6154.23 & 2.10 & $-1.58$ & 43.8\nl
Na I & 6160.75 & 2.10 & $-1.26$ & 63.1\nl
Mg I & 5711.10 & 4.34 & $-1.71$ & 110.3\nl
Si I & 6125.03 & 5.61 & $-1.54$ & 47.3\nl
Si I & 6145.02 & 5.61 & $-1.42$ & 54.3\nl
Si I & 6721.84 & 5.86 & $-1.14$ & 71.6\nl
S I & 6052.68 & 7.87 & $-0.44$ & 19.6\nl
Ca I & 5867.57 & 2.93 & $-1.62$ & 35.0\nl
Ca I & 6166.44 & 2.52 & $-1.13$ & 79.2\nl
Sc II & 5526.82 & 1.77 & $+0.10$ & 96.8\nl
Sc II & 6604.60 & 1.36 & $-1.17$ & 50.6\nl
Ti I & 5965.84 & 1.88 & $-0.38$ & 33.3\nl
Ti I & 6126.22 & 1.07 & $-1.41$ & 21.3\nl
Ti I & 6261.11 & 1.43 & $-0.46$ & 48.3\nl
Ti II & 5336.79 & 1.58 & $-1.61$ & 90.6\nl
Ti II & 5418.78 & 1.58 & $-2.07$ & 67.6\nl
Cr I & 5787.93 & 3.32 & $-0.11$ & 52.2\nl
Fe I & 5044.22 & 2.85 & $-2.04$ & 80.7\nl
Fe I & 5247.06 & 0.09 & $-4.93$ & 63.3\nl
Fe I & 5322.05 & 2.28 & $-2.86$ & 65.5\nl
Fe I & 5806.73 & 4.61 & $-0.90$ & 64.5\nl
Fe I & 5852.23 & 4.55 & $-1.18$ & 47.6\nl
Fe I & 5855.09 & 4.61 & $-1.52$ & 27.2\nl
Fe I & 5856.10 & 4.29 & $-1.56$ & 39.4\nl
Fe I & 5956.71 & 0.86 & $-4.55$ & 58.0\nl
Fe I & 6027.06 & 4.08 & $-1.09$ & 74.0\nl
Fe I & 6034.04 & 4.31 & $-2.26$ & 10.3\nl
Fe I & 6054.08 & 4.37 & $-2.20$ & 11.5\nl
Fe I & 6056.01 & 4.73 & $-0.40$ & 85.4\nl
Fe I & 6079.02 & 4.65 & $-1.02$ & 55.2\nl
Fe I & 6089.57 & 5.02 & $-0.86$ & 38.3\nl
Fe I & 6151.62 & 2.18 & $-3.29$ & 50.4\nl
Fe I & 6157.73 & 4.07 & $-1.25$ & 75.0\nl
Fe I & 6159.38 & 4.61 & $-1.87$ & 16.3\nl
Fe I & 6165.36 & 4.14 & $-1.47$ & 53.7\nl
Fe I & 6180.21 & 2.73 & $-2.61$ & 65.4\nl
Fe I & 6187.99 & 3.94 & $-1.61$ & 52.3\nl
Fe I & 6200.32 & 2.61 & $-2.44$ & 83.6\nl
Fe I & 6226.74 & 3.88 & $-2.03$ & 35.8\nl
Fe I & 6229.23 & 2.84 & $-2.82$ & 42.8\nl
Fe I & 6240.65 & 2.22 & $-3.32$ & 51.8\nl
Fe I & 6265.14 & 2.18 & $-2.57$ & 91.0\nl
Fe I & 6270.23 & 2.86 & $-2.57$ & 59.4\nl
Fe I & 6303.46 & 4.32 & $-2.55$ & 6.2\nl
Fe I & 6380.75 & 4.19 & $-1.32$ & 60.1\nl
Fe I & 6385.73 & 4.73 & $-1.82$ & 15.1\nl
Fe I & 6392.54 & 2.28 & $-4.01$ & 19.5\nl
Fe I & 6498.95 & 0.96 & $-4.62$ & 44.6\nl
Fe I & 6581.22 & 1.48 & $-4.66$ & 17.6\nl
Fe I & 6591.33 & 4.59 & $-1.98$ & 14.2\nl
Fe I & 6608.04 & 2.28 & $-4.00$ & 16.9\nl
Fe I & 6627.56 & 4.55 & $-1.44$ & 34.9\nl
Fe I & 6646.97 & 2.61 & $-3.85$ & 9.8\nl
Fe I & 6653.91 & 4.15 & $-2.41$ & 13.8\nl
Fe I & 6703.58 & 2.76 & $-3.01$ & 38.7\nl
Fe I & 6710.32 & 1.48 & $-4.80$ & 14.3\nl
Fe I & 6725.36 & 4.10 & $-2.18$ & 20.3\nl
Fe I & 6726.67 & 4.61 & $-1.04$ & 53.6\nl
Fe I & 6733.15 & 4.64 & $-1.45$ & 31.9\nl
Fe I & 6739.52 & 1.56 & $-4.90$ & 10.7\nl
Fe I & 6745.11 & 4.58 & $-2.06$ & 10.0\nl
Fe I & 6745.98 & 4.07 & $-2.68$ & 7.9\nl
Fe I & 6746.98 & 2.61 & $-4.41$ & 2.6\nl
Fe I & 6750.16 & 2.42 & $-2.62$ & 76.4\nl
Fe I & 6752.72 & 4.64 & $-1.20$ & 45.0\nl
Fe I & 6786.86 & 4.19 & $-1.95$ & 35.3\nl
Fe I & 6839.84 & 2.56 & $-3.36$ & 34.8\nl
Fe I & 6855.72 & 4.61 & $-1.73$ & 25.1\nl
Fe I & 6861.95 & 2.42 & $-3.80$ & 19.5\nl
Fe I & 6862.50 & 4.56 & $-1.35$ & 35.3\nl
Fe I & 6864.32 & 4.56 & $-2.30$ & 9.7\nl
Fe I & 7498.54 & 4.14 & $-2.09$ & 23.8\nl
Fe I & 7507.27 & 4.41 & $-1.05$ & 67.6\nl
Fe II & 5234.63 & 3.22 & $-2.20$ & 109.5\nl
Fe II & 6084.11 & 3.20 & $-3.75$ & 36.3\nl
Fe II & 6149.25 & 3.89 & $-2.70$ & 60.9\nl
Fe II & 6247.56 & 3.89 & $-2.30$ & 78.6\nl
Fe II & 6369.46 & 2.89 & $-4.11$ & 33.0\nl
Fe II & 6416.93 & 3.89 & $-2.60$ & 61.9\nl
Fe II & 6432.68 & 2.89 & $-3.29$ & 65.6\nl
Fe II & 7515.84 & 3.90 & $-3.36$ & 27.4\nl
Ni I & 6767.78 & 1.83 & $-2.09$ & 84.6\nl
Zn I & 4722.16 & 4.03 & $-0.26$ & 75.7\nl
\enddata
\end{deluxetable}

\clearpage

\begin{deluxetable}{lcccc}
\tablecaption{Equivalent Widths for $\upsilon$\,And, $\tau$\,Boo, 
and $\rho^{\rm 1}$\,Cnc}
\tablewidth{0pt}
\tablehead{
\colhead{Species} & \colhead{$\lambda_{\rm o}$(\AA)} & 
\colhead{$\upsilon$\,And} & 
\colhead{$\tau$\,Boo} & \colhead{$\rho^{\rm 1}$\,Cnc}
}
\startdata
C I & 5380.32 & 46.4 & 35.0 & 21.0\nl
C I & 6587.62 & 35.6 & 48.5 & \nodata\nl
C I & 7108.92 & 13.7 & 22.8 & \nodata\nl
C I & 7115.17 & 40.7 & 53.0 & \nodata\nl
C I & 7116.96 & \nodata & 52.6 & \nodata\nl
C I & 7483.42 & 18.3 & 24.5 & \nodata\nl
N I & 7468.27 & 9.2 & 16.7 & \nodata\nl
Na I & 6154.23 & 32.8 & 34.6 & 96.0\nl
Na I & 6160.75 & 53.6 & 58.2 & 109.2\nl
Mg I & 5711.10 & 96.8 & 98.8 & \nodata\nl
Si I & 5793.08 & 49.2 & 56.8 & 63.0\nl
Si I & 6125.03 & 36.0 & 43.5 & 52.2\nl
Si I & 6145.02 & 45.9 & 49.1 & 58.0\nl
Si I & 6721.84 & 51.8 & 64.6 & \nodata\nl
S I & 6052.68 & 25.0 & 38.2 & \nodata\nl
Ca I & 5867.57 & 23.5 & 21.6 & 57.9\nl
Ca I & 6166.44 & 64.2 & 64.6 & 111.9\nl
Sc II & 5526.82 & \nodata & \nodata & 86.9\nl
Sc II & 6604.60 & 46.5 & 44.8 & 56.0\nl
Ti I & 5965.84 & 22.1 & 19.8 & \nodata\nl
Ti I & 6126.22 & 14.3 & 13.4 & 67.0\nl
Ti I & 6261.11 & 36.9 & 42.1 & \nodata\nl
Ti II & 5336.79 & 92.2 & 97.0 & 79.5\nl
Ti II & 5418.78 & 65.1 & 78.1 & 57.8\nl
Cr I & 5787.93 & 41.3 & 48.8 & 84.4\nl
Fe I & 5044.22 & 70.8 & 67.1 & \nodata\nl
Fe I & 5247.06 & \nodata & \nodata & 100.0\nl
Fe I & 5806.73 & 51.0 & 56.4 & \nodata\nl
Fe I & 5852.23 & 44.7 & 38.3 & 71.5\nl
Fe I & 5855.09 & 21.1 & 20.7 & 46.0\nl
Fe I & 5856.10 & 27.6 & 31.4 & 60.5\nl
Fe I & 5956.71 & 36.5 & 29.6 & 84.5\nl
Fe I & 6027.06 & 63.2 & 68.1 & \nodata\nl
Fe I & 6034.04 & \nodata & \nodata & 23.6\nl
Fe I & 6054.08 & \nodata & \nodata & 28.6\nl
Fe I & 6056.01 & 73.5 & 76.0 & \nodata\nl
Fe I & 6065.48 & 111.0 & 110.1 & \nodata\nl
Fe I & 6079.02 & \nodata & \nodata & 79.0\nl
Fe I & 6089.57 & 32.5 & 37.9 & 60.6\nl
Fe I & 6093.65 & \nodata & \nodata & 56.1\nl
Fe I & 6096.67 & \nodata & \nodata & 69.8\nl
Fe I & 6098.25 & \nodata & \nodata & 38.5\nl
Fe I & 6151.62 & 37.6 & 33.6 & 81.2\nl
Fe I & 6157.73 & \nodata & \nodata & 98.0\nl
Fe I & 6159.38 & \nodata & \nodata & 35.1\nl
Fe I & 6165.36 & 43.0 & 42.2 & 67.0\nl
Fe I & 6180.21 & \nodata & \nodata & 94.7\nl
Fe I & 6187.99 & \nodata & \nodata & 83.5\nl
Fe I & 6200.32 & 68.5 & 64.9 & \nodata\nl
Fe I & 6226.74 & \nodata & \nodata & 58.2\nl
Fe I & 6229.23 & \nodata & \nodata & 80.6\nl
Fe I & 6240.65 & \nodata & \nodata & 81.6\nl
Fe I & 6270.23 & \nodata & \nodata & 86.3\nl
Fe I & 6303.46 & \nodata & \nodata & 14.7\nl
Fe I & 6380.75 & 50.0 & 53.7 & 85.9\nl
Fe I & 6385.73 & \nodata & \nodata & 28.6\nl
Fe I & 6392.59 & \nodata & \nodata & 49.9\nl
Fe I & 6498.95 & 34.8 & 28.5 & 91.3\nl
Fe I & 6581.22 & \nodata & \nodata & 61.3\nl
Fe I & 6591.33 & 8.9 & 8.5 & 27.2\nl
Fe I & 6608.04 & 13.3 & 14.4 & 49.6\nl
Fe I & 6627.56 & \nodata & \nodata & 56.8\nl
Fe I & 6646.97 & \nodata & \nodata & 33.6\nl
Fe I & 6653.91 & \nodata & \nodata & 29.7\nl
Fe I & 6703.58 & 28.7 & 26.0 & 70.8\nl
Fe I & 6710.32 & \nodata & \nodata & 56.7\nl
Fe I & 6725.36 & \nodata & \nodata & 43.8\nl
Fe I & 6726.67 & \nodata & \nodata & 77.9\nl
Fe I & 6733.15 & \nodata & \nodata & 52.6\nl
Fe I & 6739.52 & 6.4 & 7.3 & 40.6\nl
Fe I & 6745.11 & \nodata & \nodata & 24.7\nl
Fe I & 6745.98 & \nodata & \nodata & 29.1\nl
Fe I & 6746.98 & \nodata & \nodata & 18.7\nl
Fe I & 6750.16 & 68.5 & 63.6 & \nodata\nl
Fe I & 6752.72 & 33.8 & 33.0 & 69.7\nl
Fe I & 6786.86 & \nodata & \nodata & 52.9\nl
Fe I & 6820.37 & 36.2 & 39.6 & 71.3\nl
Fe I & 6833.25 & 8.6 & 10.1 & \nodata\nl
Fe I & 6839.84 & 32.4 & 31.4 & \nodata\nl
Fe I & 6855.72 & \nodata & \nodata & 42.3\nl
Fe I & 6861.95 & \nodata & \nodata & 56.5\nl
Fe I & 6862.50 & \nodata & \nodata & 59.5\nl
Fe I & 6864.32 & \nodata & \nodata & 19.7\nl
Fe I & 7498.54 & 15.8 & 17.2 & \nodata\nl
Fe I & 7507.27 & 53.3 & 55.0 & \nodata\nl
Fe I & 7583.80 & 78.2 & 76.4 & \nodata\nl
Fe I & 7586.03 & 111.3 & 114.8 & \nodata\nl
Fe II & 5234.63 & \nodata & \nodata & 85.5\nl
Fe II & 5991.38 & 50.3 & 59.3 & 32.9\nl
Fe II & 6084.11 & \nodata & \nodata & 22.0\nl
Fe II & 6149.25 & 57.3 & 64.9 & 33.4\nl
Fe II & 6247.56 & \nodata & \nodata & 44.0\nl
Fe II & 6369.46 & 31.8 & 34.8 & 18.6\nl
Fe II & 6416.93 & \nodata & \nodata & 49.3\nl
Fe II & 6432.68 & \nodata & \nodata & 40.1\nl
Fe II & 6442.95 & \nodata & \nodata & 7.2\nl
Fe II & 6446.40 & \nodata & \nodata & 3.7\nl
Fe II & 7515.84 & 25.9 & 35.4 & \nodata\nl
Ni I & 6767.78 & 72.8 & 69.9 & 115.8\nl
Zn I & 4722.16 & 80.4 & 76.6 & \nodata\nl
\enddata
\end{deluxetable}

\clearpage

\begin{deluxetable}{lcccc}
\tablecaption{Equivalent Widths for HD\,187123, HD\,210277, 
and 14\,Her}
\tablewidth{0pt}
\tablehead{
\colhead{Species} & \colhead{$\lambda_{\rm o}$(\AA)} & \colhead{HD\,187123} & 
\colhead{HD\,210277} & \colhead{14\,Her}
}
\startdata
C I & 5380.32 & 26.6 & 25.3 & 23.9\nl
Na I & 6154.23 & 45.9 & 65.1 & 99.5\nl
Na I & 6160.75 & 62.6 & 81.4 & 114.6\nl
Mg I & 5711.10 & 113.0 & 138.0 & 179.6\nl
Si I & 6125.03 & 39.0 & 46.2 & 55.9\nl
Si I & 6145.02 & 46.2 & 54.5 & 60.1\nl
Si I & 6721.84 & 57.4 & 64.1 & 80.5\nl
S I & 6052.68 & 15.0 & 17.2 & \nodata\nl
Ca I & 6166.44 & 77.5 & 94.1 & 116.6\nl
Sc II & 6604.60 & 40.5 & 45.4 & 54.7\nl
Ti I & 6126.22 & 25.8 & 47.0 & 69.2\nl
Ti II & 5336.79 & 79.4 & 87.5 & 69.2\nl
Ti II & 5418.78 & 54.1 & 55.9 & 58.9\nl
Cr I & 5787.93 & 50.3 & 64.8 & 84.0\nl
Ni I & 6767.78 & 81.7 & 95.6 & 120.6\nl
\enddata
\end{deluxetable}

\clearpage

\begin{deluxetable}{lccccccc}
\tablecaption{Spectroscopically-Determined Physical Parameters of 
$\upsilon$\,And, $\tau$\,Boo, and HD\,75289}
\tablewidth{0pt}
\tablehead{
\colhead{Star} & \colhead{T$_{\rm eff}$} & \colhead{$\log g$} & 
\colhead{$\xi_{\rm t}$} & \colhead{[Fe/H]} & 
\colhead{M$_{\rm V}$\tablenotemark{a}} & 
\colhead{Age\tablenotemark{b}} & \colhead{Mass\tablenotemark{b}} \\
\colhead{} & \colhead{(K)} & \colhead{} & 
\colhead{(km~s$^{\rm -1}$)} & \colhead{} & \colhead{} & 
\colhead{(Gyr)} & \colhead{(M$_{\odot}$)}
}
\startdata
$\upsilon$\,And & $6140 \pm 60$ & $4.12 \pm 0.11$ & $1.35 \pm 0.10$ & 
$0.12 \pm 0.05$ & $3.45 \pm 0.03$ & $3.3 \pm 0.5$ & $1.28 \pm 0.02$ \nl
$\tau$\,Boo & $6420 \pm 80$ & $4.18 \pm 0.08$ & $1.25 \pm 0.11$ & 
$0.32 \pm 0.06$ & $3.53 \pm 0.03$ & $1.5 \pm 0.5$ & $1.34 \pm 0.02$ \nl
HD\,75289 & $6140 \pm 50$ & $4.47 \pm 0.24$ & $1.48 \pm 0.10$ & 
$0.28 \pm 0.05$ & $4.04 \pm 0.04$ & $2.1^{\rm +0.70}_{\rm -0.60}$ 
& $1.22 \pm 0.02$ \nl
\enddata
\tablenotetext{a}{Calculated from the {\it Hipparcos} parallaxes.}
\tablenotetext{b}{Derived from Schaller et al. (1992) and Schaerer et al. 
(1993) stellar evolutionary isochrones.}
\end{deluxetable}

\clearpage

\begin{deluxetable}{lccccc}
\tablecaption{[X/H] values for $\upsilon$\,And, $\tau$\,Boo, 
$\rho^{\rm 1}$\,Cnc, and HD\,75289}
\tablewidth{0pt}
\tablehead{
\colhead{Element} & \colhead{$\log \epsilon_{\odot}$} & 
\colhead{$\upsilon$\,And} & \colhead{$\tau$\,Boo} & 
\colhead{$\rho^{\rm 1}$\,Cnc\tablenotemark{a}} & 
\colhead{HD\,75289}
}
\startdata
Li  & 1.06 & $+1.20 \pm 0.07$ & $+0.62 \pm 0.25$ & $<-0.60 \pm 0.15$ & 
$+1.70 \pm 0.05$\nl
C & 8.56 & $+0.06 \pm 0.11$ & $+0.06 \pm 0.09$ & $+0.33 \pm 0.10$ & 
$+0.16 \pm 0.10$\nl
N & 8.05 & $-0.10 \pm 0.09$ & $+0.04 \pm 0.10$ & $+0.71 \pm 0.14$ & 
$+0.05 \pm 0.10$\nl
Na & 6.33 & $+0.09 \pm 0.07$ & $+0.27 \pm 0.09$ & $+0.34 \pm 0.09$ & 
$+0.20 \pm 0.05$\nl
Mg & 7.58 & $+0.04 \pm 0.08$ & $+0.24 \pm 0.09$ & \nodata & 
$+0.13 \pm 0.09$\nl
Al & 6.47 & $+0.02 \pm 0.06$ & $+0.15 \pm 0.10$ & $+0.50 \pm 0.06$ & 
\nodata\nl
Si & 7.55 & $+0.16 \pm 0.02$ & $+0.38 \pm 0.04$ & $+0.39 \pm 0.02$ & 
$+0.33 \pm 0.04$\nl
S & 7.21 & $+0.11 \pm 0.08$ & $+0.29 \pm 0.08$ & \nodata & $+0.04 \pm 0.09$\nl
Ca & 6.36 & $+0.14 \pm 0.08$ & $+0.28 \pm 0.09$ & $+0.31 \pm 0.10$ & 
$+0.33 \pm 0.08$\nl
Sc & 3.10 & $+0.07 \pm 0.08$ & $+0.15 \pm 0.09$ & $+0.56 \pm 0.09$ & 
$+0.32 \pm 0.10$\nl
Ti I & 4.99 & $+0.06 \pm 0.06$ & $+0.30 \pm 0.08$ & $+0.33 \pm 0.11$ & 
$+0.26 \pm 0.05$\nl
Ti II & 4.99 & $+0.16 \pm 0.09$ & $+0.47 \pm 0.10$ & $+0.40 \pm 0.10$ & 
$+0.29 \pm 0.10$\nl
Cr & 5.67 & $+0.10 \pm 0.08$ & $+0.42 \pm 0.09$ & $+0.34 \pm 0.10$ & 
$+0.25 \pm 0.07$\nl
Fe & 7.47 & $+0.12 \pm 0.05$ & $+0.32 \pm 0.06$ & $+0.45 \pm 0.05$ & 
$+0.28 \pm 0.05$\nl
Ni & 6.25 & $+0.02 \pm 0.09$ & $+0.24 \pm 0.11$ & $+0.44 \pm 0.09$ & 
$+0.16 \pm 0.09$\nl
Zn & 4.60 & $+0.10 \pm 0.09$ & $+0.22 \pm 0.11$ & \nodata & $-0.02 \pm 0.08$\nl
\enddata
\tablenotetext{a}{The values of [Li/H], [C/H], and [N/H] for 
$\rho^{\rm 1}$\,Cnc are from Paper III.  The [C/H] value listed here is based 
on the C I line at 5380 \AA.}
\end{deluxetable}

\clearpage

\begin{deluxetable}{lccc}
\tablecaption{[X/H] values for 14\,Her, HD\,187123, and HD\,210277}
\tablewidth{0pt}
\tablehead{
\colhead{Element} & \colhead{14\,Her} & 
\colhead{HD\,187123} & \colhead{HD\,210277}
}
\startdata
Li & $<-0.36$ & $+0.14 \pm 0.20$ & $<-0.26$\nl
C & $+0.36 \pm 0.12$ & $+0.10 \pm 0.05$ & $+0.24 \pm 0.08$\nl
Na & $+0.46 \pm 0.11$ & $+0.08 \pm 0.04$ & $+0.18 \pm 0.07$\nl
Mg & $+0.52 \pm 0.11$ & $+0.06 \pm 0.05$ & $+0.20 \pm 0.08$\nl
Al & $+0.43 \pm 0.07$ & $+0.17 \pm 0.06$ & $+0.38 \pm 0.07$\nl
Si & $+0.47 \pm 0.06$ & $+0.14 \pm 0.02$ & $+0.24 \pm 0.02$\nl
S & \nodata & $+0.04 \pm 0.04$ & $+0.27 \pm 0.07$\nl
Ca & $+0.43 \pm 0.12$ & $+0.14 \pm 0.05$ & $+0.24 \pm 0.08$\nl
Sc & $+0.53 \pm 0.10$ & $+0.10 \pm 0.05$ & $+0.25 \pm 0.07$\nl
Ti I & $+0.44 \pm 0.13$ & $+0.13 \pm 0.05$ & $+0.27 \pm 0.09$\nl
Ti II & $+0.49 \pm 0.10$ & $+0.13 \pm 0.05$ & $+0.31 \pm 0.07$\nl
Cr & $+0.40 \pm 0.12$ & $+0.09 \pm 0.05$ & $+0.19 \pm 0.08$\nl
Fe & $+0.50 \pm 0.05$ & $+0.16 \pm 0.05$ & $+0.24 \pm 0.05$\nl
Ni & $+0.54 \pm 0.11$ & $+0.05 \pm 0.06$ & $+0.17 \pm 0.08$\nl
\enddata
\end{deluxetable}

\clearpage

\section*{FIGURE CAPTIONS}
 
\figcaption{The [C/Fe] values are shown as dots for 73 single F and G dwarfs 
from Gustafsson et al. (1999) and 8 single F and G dwarfs from Tomkin et al. 
(1997), with 2 stars in common between the two studies.  The Sun is shown as 
an open circle and the stars-with-planets as plus signs.  The points have 
been corrected to a common galactocentric distance of 8.8 kpc by removal of 
a small trend with galactocentric distance.  A least-squares fit to the 
Gustafsson et al. sample stars is shown as a dashed line.}
 
\figcaption{Difference between observed and calculated (with equation 1) Li 
abundances for field stars from Pasquini et al., Favata et al., and 
Randich et al. (dots).  The stars-with-planets are shown as plus signs.}
 
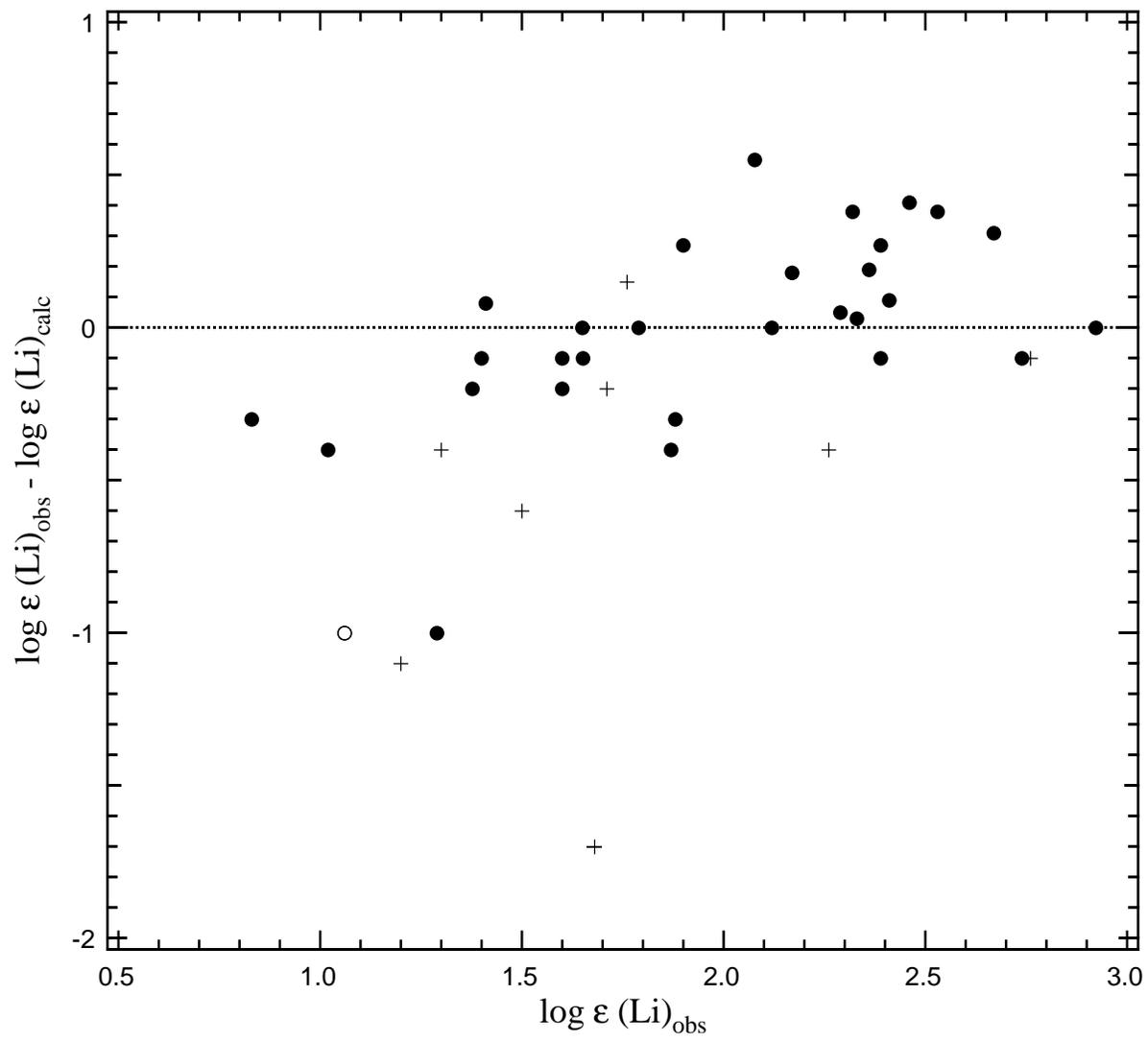
\end{document}